# A Multilayer Structure Facilitates the Production of Antifragile Systems in Boolean Network Models

Hyobin Kim[1], Omar K. Pineda[1,2] and Carlos Gershenson[1,3,4,*]

[1] Centro de Ciencias de la Complejidad, Universidad Nacional Autónoma de México, 04510 CDMX, México.

[2] Posgrado en Ciencia e Ingeniería de la Computación, Universidad Nacional Autónoma de México, 04510 CDMX, México.

[3] Instituto de Investigaciones en Matemáticas Aplicadas y en Sistemas, Universidad Nacional Autónoma de México, 04510 CDMX, México.

[4] ITMO University, St. Petersburg, 199034, Russian Federation.

* Correspondence should be addressed to Carlos Gershenson cgg@unam.mx.

## Abstract

Antifragility is a property from which systems are able to resist stress and furthermore benefit from it. Even though antifragile dynamics is found in various real-world complex systems where multiple subsystems interact with each other, the attribute has not been quantitatively explored yet in those complex systems which can be regarded as multilayer networks. Here we study how the multilayer structure affects the antifragility of the whole system. By comparing single-layer and multilayer Boolean networks based on our recently proposed antifragility measure, we found that the multilayer structure facilitated the production of antifragile systems. Our measure and findings will be useful for various applications such as exploring properties of biological systems with multilayer structures and creating more antifragile engineered systems.

## Introduction

Antifragility is a property from which systems are able to resist stress and furthermore benefit from it [1]. Although the notion of antifragility has been extensively used in many fields like computer science [2-5], transportation [6, 7], engineering [8-10], physics [11], risk analysis [12, 13], and molecular biology [14, 15], a practical quantitative measure of antifragility had not been developed. For that reason, using random Boolean networks (RBNs) and biological BNs, we recently proposed a novel metric that quantifies antifragility [16].

We measured antifragility of BNs based on the change of complexity before and after adding perturbations, in which the BNs were all single-layer networks. However, numerous real-world complex systems are composed of interacting multiple subsystems, which can be regarded as multilayer networks [17, 18]. Here we aim at investigating how the multilayer structure affects the antifragility of the whole system by assessing the antifragility of single-layer and multilayer RBNs and comparing them.

If the multilayer structure has an advantage in gaining antifragility over a single-layer structure, we could utilize the characteristic for a number of areas using BNs [19-27], from understanding properties of biological systems with multilayer structures to designing more antifragile engineered systems.

The rest of this paper is organized as follows. In the section of "Measurement of Antifragility in single-layer and multilayer RBNs", we explain single-layer and multilayer network models, how to calculate their complexity, perturbations to networks, and how to assess the antifragility. In the section "Experiments", specific experimental designs are described. In the section of "Results and Discussion", the results about the antifragility of single-layer/multilayer RBNs and a biological BN are mentioned. The last section summarizes and concludes the paper.

## Measurement of Antifragility in Single-layer & Multilayer RBNs

### Single-layer and Multilayer Random Boolean Networks

RBNs were suggested as models of gene regulatory networks (GRNs) in cells that are present in all known living organisms [28-30]. Although RBNs are highly simplified models, they can greatly explain relevant properties of life and its possibilities. Accordingly, they have been actively used in many fields such as systems biology and artificial life [31-36]. In this study, a single-layer RBN represents a GRN at a single cell level, a multilayer RBN indicates coupled GRNs at a multicellular level.

A RBN is also called $NK$ Boolean network, where $N$ is the number of nodes, and $K$ is the number of input links per node. Here self-links are allowed. In a RBN, the links are randomly arranged, and Boolean functions are randomly assigned to each node as well. Once the topology and Boolean logic rules are determined, they are maintained. Each node represents a gene. The state of a node can have either 0 (off, inhibited) or 1 (on, activated), and it is updated by the states of input nodes and corresponding Boolean functions.

A state space of a RBN is the set of all possible configurations ($2^N$) of a system including the transitions among them. In the state space, stationary configurations are attractors (point or cyclic), and the others converging into attractors are their basin of attraction. The dynamics of RBNs is divided into ordered, chaotic, or critical regimes by the structure of the state space. The ordered and chaotic regimes indicate phases. The critical regime refers to the phase transition boundary between them. $K$ can change the dynamics of RBNs systematically: ordered for $K = 1$, critical for $K = 2$, and chaotic for $K \geq 3$, on average under internal homogeneity (*i.e.*, probability of being activated or inhibited) $p = 0.5$ [37].

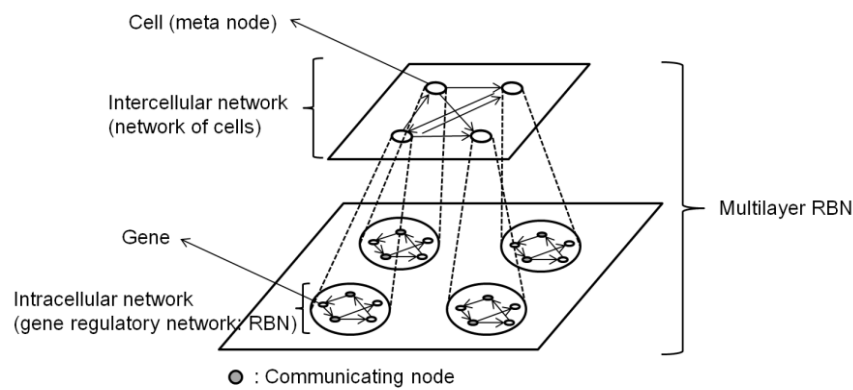

Figure 1: A schematic diagram of a multilayer RBN model. In actual simulations, the number of nodes of an intercellular network was 9, and the number of nodes of an intracellular network was 18.

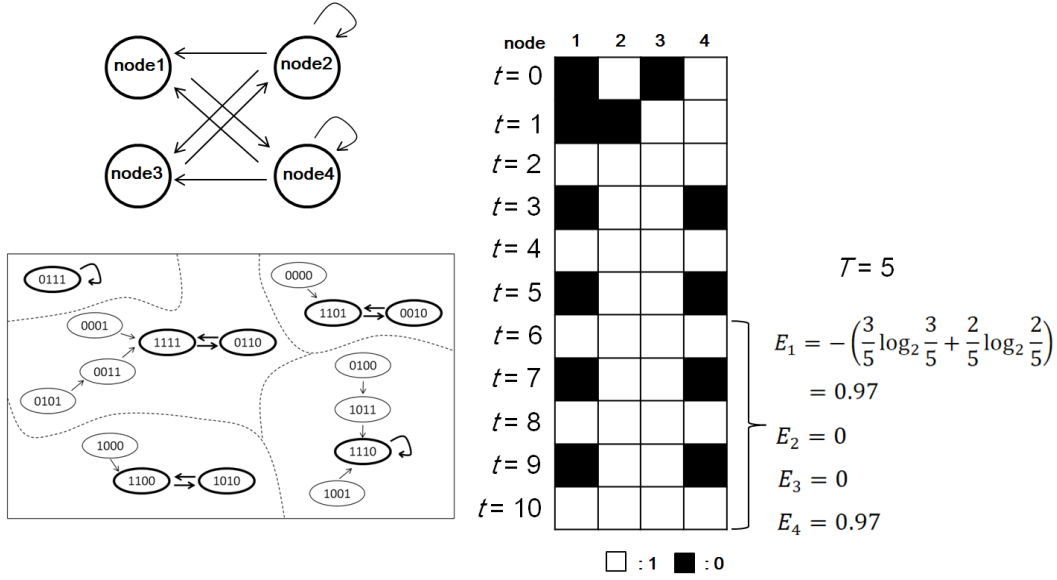

Figure 2: An example showing how to calculate complexity. Top-left: A RBN with $N = 4$, $K = 2$. Bottom-left: State space of the RBN. The state space is composed of $2^4 = 16$ configurations and transitions among them. The configurations with bold outlines are attractors. Dashed lines draw boundaries for each basin of attraction. Right: State transitions and the computation of complexity, based on the emergence of each of the four nodes (columns). 0101 was used as an initial state. The state transitions were obtained from $t = 0$ to $t = 10$.

Our multilayer RBN model is composed of two layers: intercellular and intracellular [38, 39]. In an intercellular layer, cells interact with each other and in an intracellular layer, genes interact with each other (Figure 1). All the cells have the same RBNs, and cellular topologies representing interactions between cells keep changing in each simulation. The assumption on such dynamical cellular topology is based on research showing that interacting cells continue to change by cell movements or cell growth [40].

In the multilayer RBN model, the states of all the nodes are simultaneously updated as a whole system. The specific update rules are as follows:

- Communicating genes: In each RBN, communicating nodes are assigned for cell-cell interactions, which follows cell signaling in Flann et al.'s model [31]. The state of a communicating node is determined by communicating nodes of neighboring cells. If even one is activated among the communicating nodes of its neighbors, it is activated. Here the neighbors mean source nodes from which the links originate.
- The other genes: The states of the other nodes are determined in the same way as the update of the node states in a single-layer RBN mentioned above.

**Complexity of RBNs**

We measure complexity of RBNs based on our previous approach [41, 42] as follows:

$$E_i = -(p_0 \log_2 p_0 + p_1 \log_2 p_1) \tag{1}$$

$$C = 4 \times \bar{E} \times (1 - \bar{E}) \tag{2}$$

where $E_i$ is the "emergence" of node $i$, $p_0$ ($p_1$) is the probability of how many times 0 (1) is expressed in node $i$ during $T$ time steps, $C$ ($0 \leq C \leq 1$) is the complexity of the RBN, and $\bar{E}$ ($0 \leq \bar{E} \leq 1$) is average obtained from the emergence values for every node of the network. Specifically, $p_0$ ($p_1$) is computed from simulation time $T+1$ to $2T$ not from 1 to $T$, which is to obtain $p_0$ ($p_1$) in more stable state transitions (*i.e.*, closer to attractors). Figure 2 shows an example calculating complexity of a RBN.

Emergence here means novel information, so it can be measured precisely with Shannon's information entropy (equation (1)). Complexity is conceptually understood as a balance between regularity and change [29]. In equation (2), emergence $\bar{E}$ represents change, and its complement 1-$\bar{E}$ indicates regularity. In our previous study, for regular RBNs it was maximized at the phase transition, *i.e.* in the critical regime [41]. Our complexity measure is similar to Galas et al.'s set complexity [43]. Set complexity, based on Kolmogorov's intrinsic complexity, quantifies the amount of information in a set of objects. Pairs of objects that are maximally redundant or completely random carry negligible information. They calculated the set complexity of trajectories of RBNs and found that the quantity was maximized in critical regime.

To get back to the point about our complexity measure, we can interpret the regularity and the change from an information viewpoint. Regularity enables information to be preserved, and change allows new information to be explored [42]. In the context of RBNs used as GRN models, keeping and changing the node states which point out genetic information can be connected with stability to maintain existing functions and adaptability to flexibly adapt to a new environment.

In equation (2), an optimal balance between regularity and change is achieved at $\bar{E} = 0.5$ ($\bar{E} = 0.5 \rightarrow C = 1$). That is, when either $p_0$ or $p_1$ is about 0.89, the complexity has its maximum. [41, 44]. On the contrary, the complexity becomes its minimum when the emergence $\bar{E}$ is 0 or 1 ($\bar{E} = 0$ or $1 \rightarrow C = 0$). It is when only one state is expressed ($p_0$ or $p_1 = 1$; $\bar{E} = 0$) or the two states are expressed at the same ratio ($p_0 = p_1 = 0.5$; $\bar{E} = 1$). The coefficient 4 is added to normalize $C$ to the [0,1] interval.

**Network Perturbations to RBNs**

We perturb RBNs by flipping node states. During $2 \times T$ time steps, we randomly choose $X$ nodes in a RBN consisting of $N$ nodes, and perturb the nodes with frequency $O$. The perturbations are introduced at $t \bmod O = 0$ (*i.e.*, only when the time step $t$ can be divided by $O$). For example, $X = 3$, $O = 5$, and $T = 10$ mean that we randomly choose three nodes of a network at each time step, and flip the node states every five time steps from the initial to $2 \times 10$ time steps. The perturbed three nodes are different every five time steps. Then, we calculate fragility based on the states transitions during ten time steps from $t = 11$ to $t = 20$.

To normalize fragility values between -1 and 1, we define the degree of perturbations as follows:

$$\Delta x = \frac{X \times (\frac{T}{O})}{N \times T} \qquad (3)$$

where $0 \leq \Delta x \leq 1$.

**Antifragility of RBNs**

*(anti)fragility* $\oint$ (-1 ≤ $\oint$ ≤ 1) is defined as follows:

$$\oint = -\Delta\sigma \times \Delta x \qquad (4)$$

where $\Delta\sigma$ is the difference of "satisfaction" from perturbations, and $\Delta x$ is the degree of perturbations added to a system. The satisfaction σ means the degree of how much agents attain their goal [45]. The satisfaction is contingent on what the defined system is. In this study, each node of the RBN is an agent, and their goal is defined as high complexity. In other words, networks have higher satisfaction when they are closer to criticality. The satisfaction is computed using complexity. However, one can measure the satisfaction using other criteria such as performance and fitness.

If a system does not get the satisfaction from perturbations and rather is damaged, it means the system is fragile. If the system does not change against perturbations, then it is robust. If the system increases its satisfaction with perturbations, it is antifragile.

In RBNs, the difference of satisfaction is computed based on the difference of complexity before and after adding perturbations. $\Delta\sigma$ is computed by the following equation:

$$\Delta\sigma = C - C_0 \qquad (5)$$

where $C_0$ is complexity of a RBN before perturbations are added to the network, and $C$ is complexity of the RBN after perturbations are introduced to the network. For an original RBN and its perturbed one, the same initial states are applied at $t = 0$. $C_0$ and $C$ have values between 0 and 1. Thus, $\Delta\sigma$ has values between -1 and 1 (-1 ≤ $\Delta\sigma$ ≤ 1). Regarding $\Delta x$, it was described in the section of "Network Perturbations to RBNs".

If the $\oint$ of a RBN has a negative value, the RBN is considered antifragile. If $\oint$ is a positive value, the RBN is fragile. If $\oint$ is close to zero, the RBN is robust.

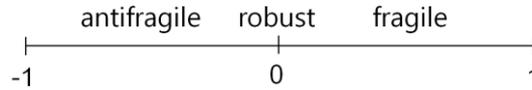

Based on equation (4), $\oint$ has negative values when $C$ is bigger than $C_0$, which means that the complexity is increased by perturbations. On the other hand, $\oint$ has positive values when $C_0$ is larger than $C$, which indicates that the complexity is reduced due to the perturbations. $\oint$ becomes zero when $C$ is equal to $C_0$. It represents that the complexity is the same before and after adding the perturbations.

## Experiments

We conducted three experiments using single-layer & multilayer RBNs and our (anti)fragility measure $\oint$ described as materials and methods in the section above.

(1) Antifragility in multilayer RBNs: We study how $\oint$ of ordered ($K = 1$), critical ($K = 2$), and chaotic ($K = 3, 4$) multilayer RBNs dynamically varies depending on the frequency and size of perturbations, and parameters related to the multilayer structure.

(2) Comparison of antifragility between multilayer and single-layer RBNs: We investigate differences between multilayer and single-layer RBNs by comparing probability of generating antifragile networks.

(3) Comparison of antifragility between multilayer and single-layer *CD4+ T-cell* networks: We examine differences between multilayer and single-layer biological BNs taking a *CD4+ T-cell* network as a real biological example. We compare average values of $\oint$ and probability of generating antifragile networks between them.

Parameter settings for simulations are shown below.

In the case of single-layer RBNs, we produced ordered, critical, and chaotic RBNs. Specifically, 10 different initial states were randomly chosen per a RBN, and then the state transitions from each initial were investigated during $2 \times 400 = 800$ time steps. Using the same initial states, we also looked into the state transitions of the perturbed RBNs. Comparing them during the last 400 time steps, we computed respective $\oint$ from the 10 initial states to obtain their mean. The plots show the values of $\oint$, which are averages from 100 different RBNs for each *K*.

For multilayer RBNs, we generated three regimes of multilayer networks taking ordered, critical, chaotic RBNs. The topology at an intercellular layer was randomly determined based on the number of links randomly chosen between 1 and 81 (because the number of cells was set to nine, the intercellular network can have a maximum of 81 links.). For an individual RBN, the genes were set to eighteen and the communicating genes were set to six, which is based on the numbers of genes and cell signaling molecules in the real biological system we used (*i.e.*, *CD4+ T-cell*). In the same manner as $\oint$ of single-layer RBNs, $\oint$ of multilayer RBNs was calculated.

Table 1: Parameters for simulations and their values

| Fig. | $N_g$ | $K$ | $N_c$ | $C_g$ | $L_i$ | $N_T$ | $T$ | $X$ | $O$ | # of different networks | # of initial states |
|---|---|---|---|---|---|---|---|---|---|---|---|
| 3(a) | 18 | 1,2,3,4 | 9 | 6 | U(1,81) | 162 | 400 | 80 | 1..50 | 100 | 10 |
| 3(b) | 18 | 1,2,3,4 | 9 | 6 | U(1,81) | 162 | 400 | 1..162 | 1 | 100 | 10 |
| 4(a) | 18 | 1,2,3,4 | 9 | 6 | U(1,81) | 162 | 400 | - | - | 100 | 10 |
| 4(b) | 18 | 1,2,3,4 | 9 | 6 | U(1,81) | 162 | 400 | 1..162 | 1 | 100 | 10 |
| 4(c) | 18 | 1,2,3,4 | 9 | 6 | U(1,81) | 162 | 400 | 1..162 | 1 | 100 | 10 |
| 5(a) | 18 | 1,2,3,4 | 9 | 1..18 | U(1,81) | 162 | 400 | 30 | 1 | 100 | 10 |
| 5(b) | 18 | 1,2,3,4 | 9 | 6 | 10..80 | 162 | 400 | 30 | 1 | 100 | 10 |
| 6(a) | 18 | 1,2,3,4 | 9 | 6 | U(1,81) | 162 | 400 | 1..162 | 1..5 | 100 | 10 |
| 6(b) | 162 | 1,2,3,4 | 1 | - | - | 162 | 400 | 1..162 | 1..5 | 100 | 10 |
| 6(c) | 18 | 1,2,3,4 | 1 | - | - | 18 | 400 | 1..18 | 1..5 | 100 | 10 |
| 8(a) | 18 | 1,2,3,4 | 9 | 6 | U(1,81) | 162 | 400 | 1..162 | 1..5 | 1 | 1000 |
| 8(b) | 18 | 1,2,3,4 | 1 | - | - | 18 | 400 | 1..18 | 1..5 | 1 | 1000 |

For a biological BN, we made use of a *CD4+ T cell* network consisting of 18 nodes. The network for *CD4+ T cell differentiation and plasticity* is modeled to study immune responses controlled by *CD4+ T cells* in terms of factors such as immunological challenges and environmental signals [46]. In the network, there are six cytokines which are cell signaling molecules related to cell-cell communications. We considered the six cytokines communicating nodes in our multilayer network model. For both a single-layer *CD4+ T cell* network and a multilayer *CD4+ T-cell* network, 1000 different initial states were randomly chosen and then the state transitions from each initial were examined during $2 \times 400 = 800$ time steps. Changing the parameters $X$ and $O$, we computed $\oint$.

For the simulation, the following parameters were used:

- Number of genes ($N_g$)
- Number of in-degrees per node ($K$)
- Number of cells ($N_c$)
- Number of communicating genes ($C_g$)
- Number of links of an intercellular network ($L_i$)
- Number of total genes at a multicellular level ($N_T = N_g \times N_c$)
- Simulation time ($T$)
- Number of perturbed genes ($X$)
- Perturbation frequency ($O$)
- Number of different networks
- Number of initial states

The specific values of parameters follow Table 1. Our simulator was implemented in Java.

## Results and Discussion

### Antifragility in Multilayer RBNs

Figure 3(a) shows average fragility of multilayer RBNs for $K = 1, 2, 3, 4$ depending on perturbation frequency $O$ (*i.e.*, the period of adding perturbations) when perturbed node size $X = 80$. With $O$ growing, the antifragile or fragile dynamics of the ordered, critical and chaotic networks changed into robust beyond $O = 30$ even though about half nodes were perturbed ($X = 80$). This result means that the perturbation frequency is more important than the perturbed node size.

Figure 3(b) represents average fragility of multilayer RBNs depending on $X$ when $O = 1$. For all $K$, there were certain ranges of $X$ for the networks to be antifragile. We also found that the certain ranges of $X$ decreased and antifragility declined as $K$ increased. These findings indicate that "optimal" antifragility results from a moderate level of perturbations, and multilayer RBNs take bigger benefits from perturbations in the order of ordered, critical, and chaotic networks. From Figure 3(a) and 3(b), we can see that maximal antifragility is obtained from a moderate level of perturbations. In addition, it is worth noting that the maximum antifragility varies for different $K$ values.

Figure 4 explains the reason why multilayer RBNs obtained more improved antifragility in the order of ordered, critical, chaotic networks. In Figure 4(a), the complexity before perturbations gradually increased as $K$ got bigger, while in Figure 4(b) the complexity after perturbations increased as $K$ got smaller excluding the early range of $X$ ($1 \leq X \leq 20$). Figure 4(c) shows the difference of complexity before and after perturbations. As seen in the figure, the smaller $K$ was, the larger the difference was. It means that the complexity representing the balance between regularity and change can be improved by perturbations, and the degree of the improvement is much larger for smaller $K$.

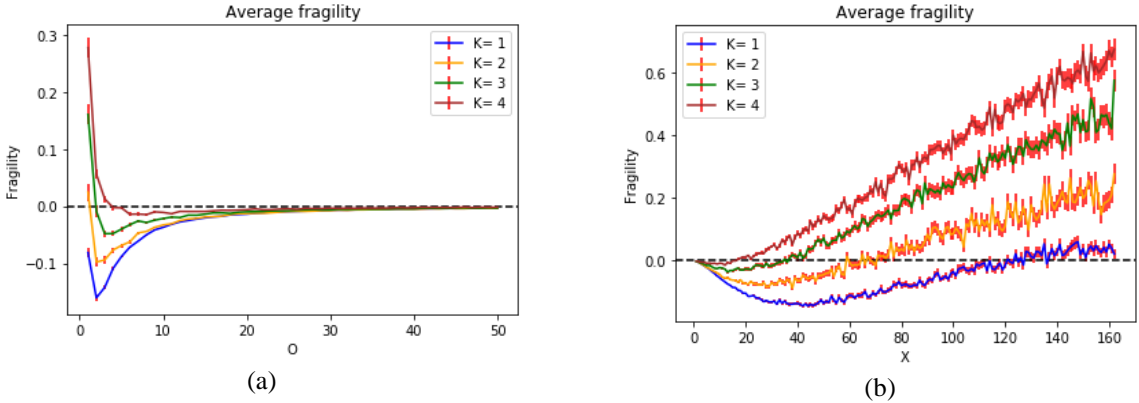

(a)   (b)

Figure 3: Average fragility of ordered ($K = 1$), critical ($K = 2$), and chaotic ($K = 3, 4$) multilayer RBNs depending on (a) $O$ and (b) $X$. The error bars represent the standard errors of measurements. For each $K$, 100 different networks were used. For each network, 10 initial states were randomly chosen. (a) $N_T = 162$, $T = 400$, and $X = 80$. (b) $N_T = 162$, $T = 400$, and $O = 1$.

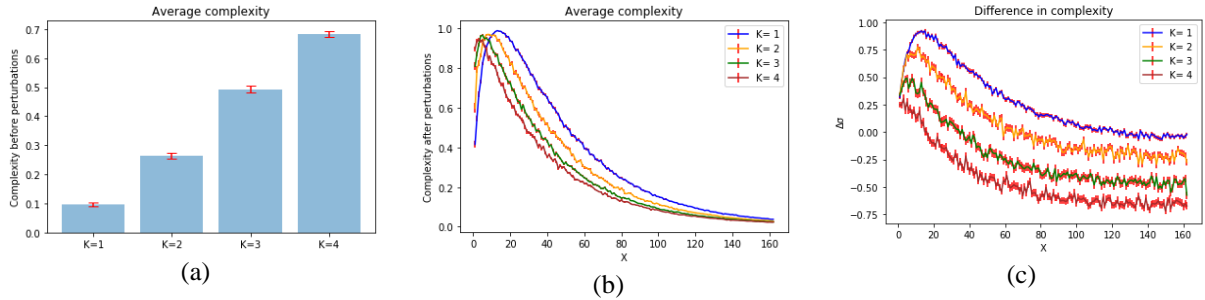

(a)   (b)   (c)

Figure 4: Average complexity of ordered, critical, and chaotic multilayer RBNs. (a) Complexity before perturbations. (b) Complexity after perturbations. (c) Difference of complexity before and after perturbations

Regarding the complexity, one interesting finding is that the complexity before perturbations (Figure 4(a)) is not consistent with that of previous studies [41, 47, 48]. The existing studies demonstrated that critical RBNs have higher complexity, while our result revealed that chaotic multilayer RBNs have larger complexity. The difference is due to a cellular level. The previous studies were performed at a single cell level, and our study was conducted at a multicellular level. The different results between multilayer and single-layer RBNs emphasize the need for research in the context of multicellular settings.

Figure 5 shows average fragility of multilayer RBNs depending on two parameters related to the multilayer structure: the number of communicating genes and the number of links of an intercellular network. In Figure 5(a), as the number of communicating genes increased, multilayer RBNs in all the regimes became antifragile. In Figure 5(b), the fragility values did not change significantly except for the early range as the number of links of an intercellular network increased. These results indicate that the communicating genes have a larger effect on antifragility at a multicellular level. Also, it suggests the possibility that the number of communicating genes representing the degree of interactions between cells might be able to be used as an indicator to estimate the effect of multilayer structure on antifragility.

**Comparison of Antifragility Between Multilayer and Single-layer RBNs**

To study how the multilayer structure has an effect on the production of antifragile networks, we calculated probability of how many antifragile networks were generated in multilayer and single-layer RBNs, respectively and then compared them. Figure 6 shows heat maps representing the probability values in a diverse range of $X$ and $O$ for multilayer and single-layer networks.

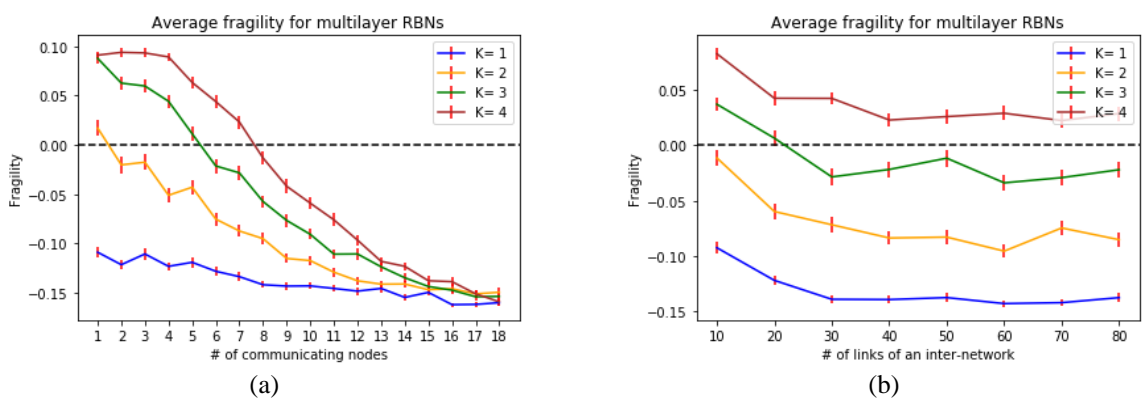

Figure 5: Average fragility of ordered, critical, and chaotic multilayer RBNs depending on two parameters related to the multilayer structure. (a) Fragility against the number of communicating nodes. (b) Fragility against the number of links of an intercellular network.

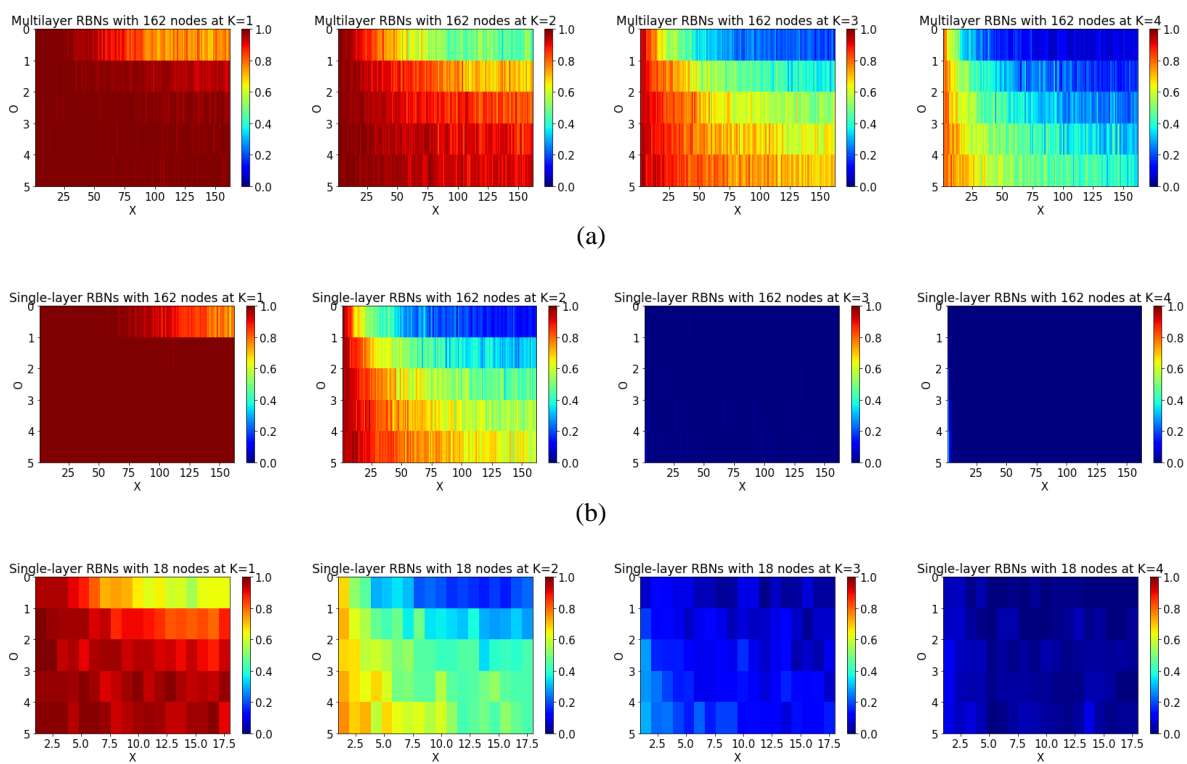

(c)

Figure 6: Probability of generating antifragile networks depending on $X$ and $O$ for $K = 1, 2, 3, 4$ (from left toward right). The probability is between 0 and 1. In the color bar, blue represents the minimum probability, and red means the maximum value. (a) Probability of producing antifragile networks in multilayer networks ($N_T = 162$). (b) Probability of producing antifragile networks in single-layer networks ($N_T = 162$) with the same state space size as multilayer networks. (c) Probability of producing antifragile networks in single-layer networks ($N_T = 18$) with the same node size as the number of genes in one cell of the multilayer network model.

Figure 6(a) is the probability in multilayer RBNs with $N_T = 162$. Figure 6(b) is the probability in single-layer RBNs with $N_T = 162$ which have the same state space size as multilayer RBNs (*i.e.*, the state space size = $2^{162}$). Figure 6(c) is the probability in single-layer RBNs with $N_T = 18$ which have the same node size as the number of genes in one cell of the multilayer network model.

For all the cases of (a), (b), and (c) in Figure 6, we found that they had similar trends that the smaller $K$ was, the more frequently antifragile networks were produced. It means that the trends of antifragile dynamics at a single-cell level are still maintained at a multicellular level. Also, as the perturbed node size increased and the period of adding perturbations became shorter, the probability of generating antifragile networks decreased overall. It is worth noticing that, especially for large $K$ values, there is not much difference between the single-layer RBNs independently of their size (Figures 6(b) and 6(c)). On the other hand, multilayer RBNs are clearly more antifragile (Figure 6(a)).

Figure 7 shows the comparison of the probability acquired from Figure 6 based on a two-sample t-test. Firstly, to investigate the difference of antifragility between multicellular and single-cell systems with the same system size, we compared multilayer RBNs with 162 nodes and single-layer RBNs with 162 nodes. As seen in the figure, the multilayer networks produced antifragile networks more frequently at $K = 2, 3, 4$. In the case of $K = 1$, the single-layer networks had higher probability, but the values were not so different.

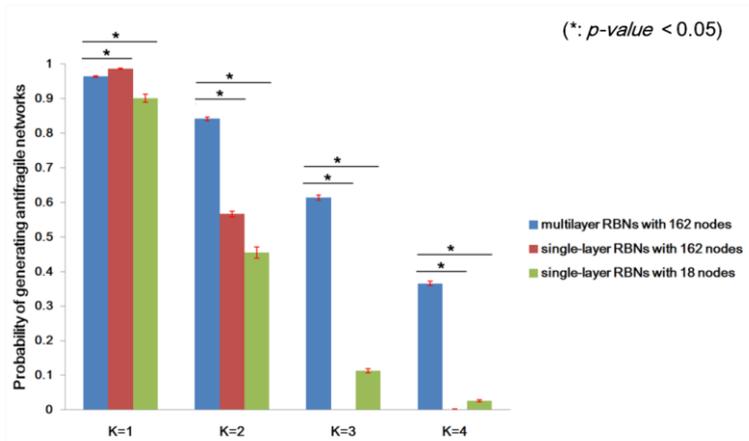

Figure 7: Comparison between multilayer and single-layer RBNs based on the probability of generating antifragile networks.

Secondly, to examine the difference of antifragility between a multicellular system and its component, we compared multilayer RBNs with 162 nodes and single-layer RBNs with 18 nodes. We found that the multilayer networks generated antifragile networks with higher probability for all $K$. The findings from Figure 7 indicate that the multilayer structure helps to produce the greater number of antifragile networks, especially for larger $K$ values.

**Comparison of Antifragility Between Multilayer and Single-layer *CD4+ T-cell* networks**

Using a *CD4+ T-cell* network related to the immune system as an example of biological systems, we calculated average fragility and the probability of generating antifragile networks for an individual *CD4+ T-cell* network and coupled *CD4+ T-cell* networks. As seen in Figure 8(a) and 8(c), the general tendency of antifragile dynamics in the *CD4+ T-cell* network at a multicellular level was practically the same as the tendency at a single-cell level.

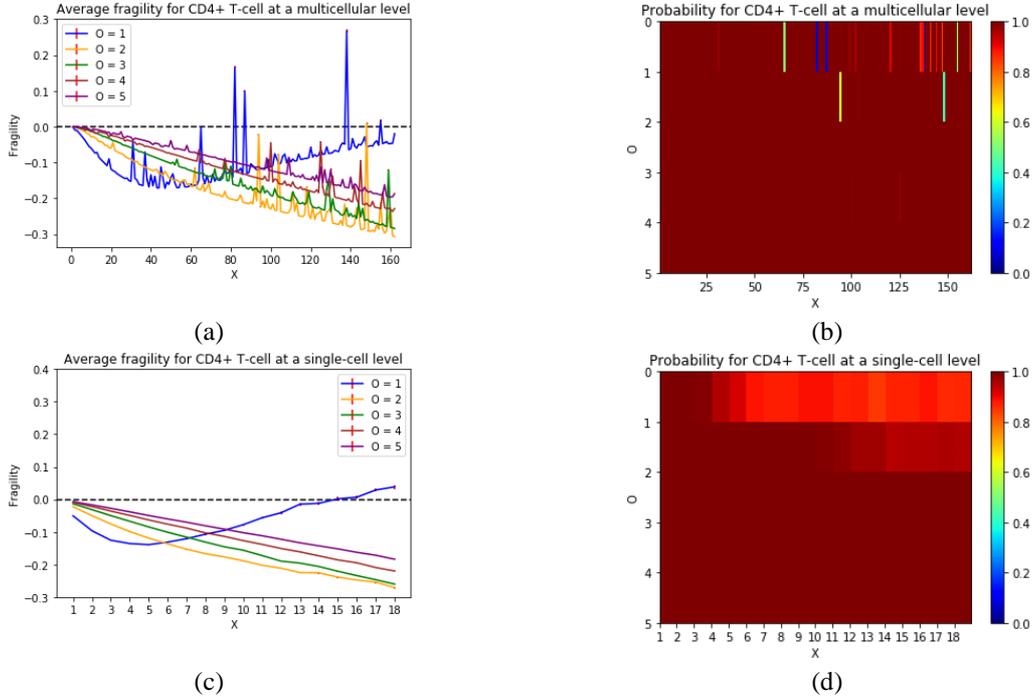

(a) (b)
(c) (d)

Figure 8: Average fragility and the probability of generating antifragile networks in multilayer and single-layer *CD4+ T-cell* networks. (a) Fragility for *CD4+ T-cell* at a multicellular level. (b) Probability for *CD4+ T-cell* at a multicellular level. (c) Fragility for *CD4+ T-cell* at a single-cell level. (d) Probability for *CD4+ T-cell* at a single-cell level

We compared multilayer and single-layer *CD4+T-cell* networks based on the probability values in Figure 8(b) and 8(d). We found that multilayer *CD4+ T-cell* networks generated antifragile networks with modestly higher probability on a two-sample t-test (*i.e.*, multicellular level: 99.26% > single-cell level: 97.74% under *p-value* < 0.05). From a biological viewpoint, these results suggest that the properties of biological systems might be enhanced in the structure of interacting multiple subsystems.

When compared to multilayer and single-layer RBNs, the *CD4+ T-cell* network showed similar antifragile dynamics to the dynamics of multilayer and single-layer networks at $K = 1$. From this, we can infer that the *CD4+ T-cell* network may be ordered. This can be understood because immune cells probably not only have a variable environment, but actually have evolved to thrive on it. The finding on the ordered dynamics of the *CD4+ T-cell* network is consistent with many research findings exhibiting that gene regulatory networks of biological systems have ordered or critical dynamics [49-52].

## Conclusions

In this study, applying our (anti)fragility measure to multilayer and single-layer BNs, we studied how the dynamics of the networks varies depending on relevant parameters, and how the multilayer structure affects the antifragility of the whole system. We found that systems showed different dynamics depending on the degree of perturbations and the degree of interaction between system components: fragile, robust or antifragile. Also, we found that the multilayer structure facilitated the production of antifragile systems. Probably this is related to the modular structure of multilayer RBNs [33], although further studies should be made.

The findings can be utilized for various applications such as systems biology and bio-inspired engineering. For example, our results may be helpful to figure out dynamical characteristics of multicellular organisms. Also, we could create engineered systems with an increased antifragility based on the fact that system properties can vary from fragile through robust to antifragile dynamics depending on the size and frequency of perturbations, and the number of communicating nodes.

Our study has a few limitations. Firstly, our multilayer RBN model is the one where identical RBNs are randomly coupled. However, there are many other systems where different subsystems are connected to each other and they communicate in a certain way. Secondly, we used only one biological example in explaining the dynamical behaviors of biological systems. To obtain more generalized findings on antifragility, we plan to develop different kinds of multilayer network models, explore them, and use various biological systems.

## Data Availability

Our source code and data are available at https://github.com/bin20005/antifragility

## Conflicts of Interest

The authors declare that there is no conflict of interest regarding the publication of this paper.

## Funding Statement

This research was partially supported by CONACYT and DGAPA, UNAM.

## Acknowledgments

We are grateful to Dario Alatorre for useful comments and discussions.